\documentclass[draft]{agujournal2019}
\usepackage{url} 
\usepackage{lineno}
\usepackage[finalnew]{trackchanges}
\usepackage{soul}
\draftfalse

\journalname{Radio Science}

\begin{document}

\title{Holographic Calibration of Phased Array Telescopes}

%
%




\authors{U. Kiefner\affil{1,2}, R. B. Wayth\affil{2}, D. B. Davidson\affil{2}, M. Sokolowski\affil{2}}

\affiliation{1}{Dept. of Electrical Engineering and Information Technology, University Paderborn, Germany}
\affiliation{2}{ICRAR/Curtin University, Perth, Australia}





\correspondingauthor{Ulrich Kiefner}{u.kiefner@posteo.de}
\correspondingauthor{Randall Wayth}{r.wayth@curtin.edu.au}


\begin{keypoints}
\item holographic aperture imaging for phased array telescopes is presented 
\item suggested technique requires few resources and can be used for calibration
\item technique was successfully tested on prototype station of SKA-Low
\end{keypoints}

%
%

%
%


\begin{abstract}
In radio astronomy, holography is a commonly used technique to create an image of the electric field distribution in the aperture of a dish antenna. The image is used to detect imperfections in the reflector surface.
Similarly, holography can be applied to phased array telescopes, in order to measure the complex gains of the receive paths of individual antennas.
In this paper, a holographic technique is suggested to calibrate the digital beamformer of a phased array telescope. The effectiveness of the technique was demonstrated by applying it on data from the Engineering Development Array 2, one of the prototype stations of the low frequency component of the Square Kilometre Array. The calibration method is very quick and requires few resources. In contrast to holography for dish antennas, it works without a reference antenna.
We demonstrate the utility of this technique for initial station commissioning and verification as well as for routine station calibration.
\end{abstract}

%
%

\section{Introduction}
Over the last decade, several new low frequency radio telescope arrays have been built where the primary receptor in the array is a dipole-like element. These arrays include the Long Wavelength Array \cite<LWA, >{2013ITAP...61.2540E}, The Precision Array for Probing the Epoch of Re-ionization \cite<PAPER, >{2010AJ....139.1468P}, the Murchison Widefield Array \cite<MWA, >{2013PASA...30....7T} and the Low Frequency Array \cite<LOFAR, >{2013A&A...556A...2V}.
These telescopes operate based on phased-array concepts, and use a combination of analogue and/or digital techniques to form the primary beam of the telescope array.

A common challenge of phased-array radio telescope systems is the characterisation and calibration of the phased-array antenna pattern, which can change significantly depending on the pointing direction on the sky. Substantial effort has been made to form predictive models of the phased-array response \cite<e.g. >{2015RaSc...50...52S,2017PASA...34...62S,2017ExA....44..239B,2018ITAP...66.1805B,Warnicketal}, which in turn requires an accurate electromagnetic model of the elements in the phased-array, including mutual coupling effects.
Work to characterise the element or primary beam pattern of the telescope includes the use of Unmanned Aerial Vehicles (UAVs) with radio transmitters, low earth orbit satellites and astronomical sources
\cite{2015ITAP...63.5433S,2015RaSc...50..614N,2016SPIE.9906E..3VB,2017PASP..129c5002J,2018PASA...35...45L}.

Having an accurate model of the primary beam, and hence primary beam aperture, is driven in large part by the demanding calibration requirements for studying the faint radio signals from the early universe \cite<e.g. >[chap 5]{2019cosm.book.....M}. The lessons and techniques learned from the existing radio telescope arrays are also directly applicable to the future Square Kilometre Array Low-frequency array radio telescope (SKA-Low).
In this paper, we present the results of using a holographic technique to directly measure the complex aperture illumination pattern for a prototype SKA-Low ``station'', which is comprised of 256 antennas working as a phased array, where signals from each antenna are individually digitised and combined digitally.
{The prototype system used in this paper is described in Section} \ref{sec:measurements}and the specifications of the SKA-Low stations can be found in \citeA{SKA1BaselineDesign}.

\section{Holographic Aperture Imaging}
Holographic aperture imaging is based on the Fourier relationship between the distribution of the electric field in the receiver's aperture and its far field electric field (voltage) reception pattern \cite[p. 819]{2017isra.book.....T}: 
\begin{equation}\label{eq:fourier}
E_\mathrm{A}(x,y) = \int_{-\infty}^{\infty}\int_{-\infty}^{\infty} V_\mathrm{F}(l,m)\cdot\mathrm{e}^{-\mathrm{j}2\pi(x l + y m)}\,dl\,dm,
\end{equation}
where {$x$} and {$y$} are the east and north coordinates, respectively, in the aperture and measured in wavelength. $l$ and $m$ are the standard radio interferometric direction cosines with their origin at the phase centre and are aligned with the celestial Right Ascention and Declination coordinates. $E_\mathrm{A}$ is the complex aperture electric field distribution and $V_\mathrm{F}$ the far field voltage reception pattern (beam pattern). By taking the two dimensional Fourier Transformation of the beam pattern, the aperture distribution can be obtained, including magnitude and phase information. The required complex beam pattern can be measured by scanning the beam of the antenna under test and making use of a second receiver, which acts as a reference. 

\subsection{Aperture Imaging for Dish Antennas}
\begin{figure}
\centering
\includegraphics[trim={1cm 4cm 1cm 4cm},clip,width=0.5\textwidth]{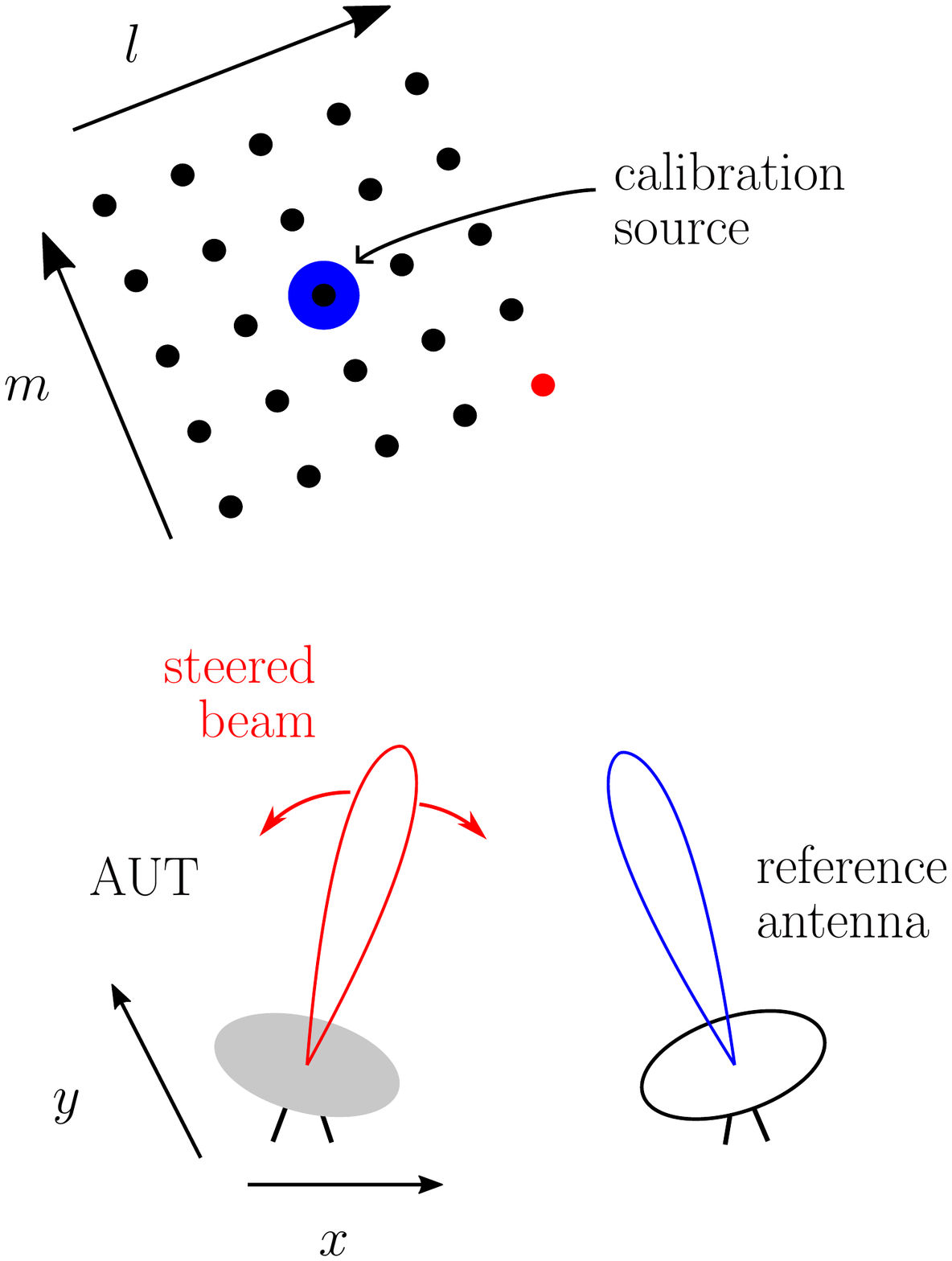}
\caption{Sketch of the holographic measurement procedure for dish antennas. A separate reference antenna stays focused on the calibration source, while the AUT is adjusted to various $lm$-coordinates. \add[RW]{The coordinate axes are drawn such that we are looking ``down'' on both the antennas and the sky.}}
\label{fig:dish}
\end{figure}

The commonly used methodology for dish antennas, which was first described by \citeA{Napier}, is illustrated in Figure \ref{fig:dish}. The beam pattern of the antenna under test (AUT) is sampled on a grid of $lm$-coordinates which is centered around an unresolved calibration source. The dish is scanned under the source, so the beam points consecutively towards all grid coordinates during the measurement. At the same time, a second reference antenna stays focused on the calibration source. By cross-correlating the received signals of AUT and reference antenna for every sampling coordinate, a complex quantity is obtained. Because it is proportional to the beam pattern of the AUT, it can be treated as a voltage-like quantity and we can use it in equation (\ref{eq:fourier}) for $V_\mathrm{F}(l,m)$. The measurements with dish antennas are usually very sophisticated, because the antennas have to be adjusted mechanically, and the calibration source changes its celestial position during the lengthy process.
\begin{figure}
\centering
\includegraphics[trim={1cm 4cm 1cm 4cm},clip,width=0.5\textwidth]{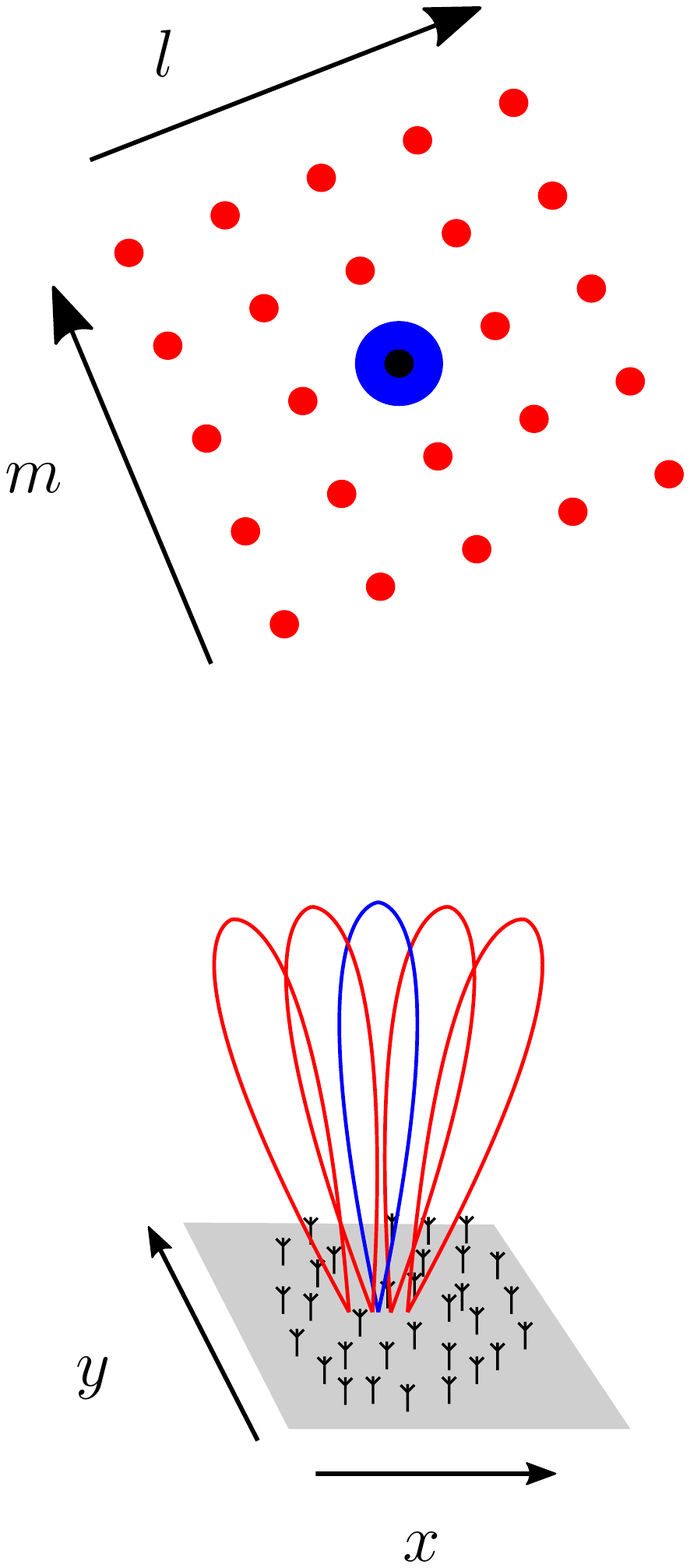}
\caption{Sketch of holographic measurements for phased arrays. All beams can be formed at once in post-processing.}\label{fig:array}
\end{figure}

\subsection{Suggested Method for Phased Arrays}\label{sec:suggested_method_arrays}
Our suggested technique for phased arrays follows the same principle. Similar approaches have been used to characterize time delays between the stations of LOFAR \cite{2020A&A...635A.207S} and also to calibrate it on station level \cite{Brentjens}. However, the method presented here is a more direct transfer of holography from dish antennas to phased array telescopes. It is assumed that the telescope under test (TUT) digitises the individual signals of all antennas and captures a short snapshot of them while a strong calibrator source is in the field of view of the telescope. Then, all required information can be gathered from these data, as beams can be formed to all coordinates on the $lm$-grid in post-processing. Furthermore, even without initial phase calibration, the TUT can form an (unprecise) beam towards the calibration source, which can be used as reference beam. Hence, a separate reference antenna is not required (see Figure \ref{fig:array}). This aspect of the technique was already proposed in \citeA{Wijnholds}. While we form cross-correlations between beamformed outputs and the reference beam signal, Wijnholds correlates individual antenna signals with the reference beam signal.


For aperture arrays where mutual coupling is minimal, the station beam can be computed using classical phased array theory: the beam is the product of the array factor and the element pattern (identical for all elements). 
By steering the beam (pointing to different $lm$-coordinates) and de-embedding the array factor, the element pattern could potentially be established from one calibrator
\change[RW]{, although the accuracy of this process may be influenced by signal-to-noise ratio considersatons and in practice, multiple calibrators are likely to be required.}{by taking many snapshots separated in time as the source moves through the sky, or by using multiple calibrators.} 

However, for arrays where mutual coupling cannot be neglected, the station beam cannot be decomposed into an array factor and element pattern. Instead, it must be computed as the (possibly weighted) sum of the individual embedded element patterns (EEPs) \cite{Warnicketal}. An EEP is the radiation pattern from an individual element in the array, with all the other elements suitably terminated. It is defined to include the phase term resulting from its position in the array, although this may be more conveniently inserted subsequently\footnote{Although similar to an array factor, these phase terms each multiply a different EEP, rather than the array factor which multiplies the element pattern.}. To measure the EEPs, all but one antenna must be terminated; as such, there is no way to scan the beam, so for a tightly coupled array, such as SKA-Low prototypes using log-period dipole elements \cite{DavidsonBolli_etal_ICEAA19}, a calibrator in a particular direction provides information on only the EEP in that direction. Either a large number of calibrators, or a moving calibrators such as a UAVs or satellites, would be required to establish the full EEPs.


The aperture image, which is the Fourier Transformation of the obtained array factor, contains information about the receive path gains and phases of all antennas and can be used for calibration of the array. 



Assume the TUT comprises  $N$ antennas, which are positioned at the coordinates {$(x_n,y_n)$} around the origin of the array. The coordinates are normalised to the observed wavelength.
Antenna $n$ receives the signal $f_n[k]$ , where {$k\in \{1,\cdots,K\}$} is the sample index. The calibration source is located at the celestial coordinates $(l_\mathrm{ref},m_\mathrm{ref})$.
By applying the appropriate phase factors $a_n$ to the antenna signals\footnote{This process requires prior knowledge of the locations of antennas and mappings between antenna and digital signals, however the technique also has utility in verifying these mappings as discussed in Section \ref{sec:measurements}.},
a beam is formed towards the calibration source and the reference signal {$g_{\mathrm{ref}}[k]$} can be computed:
\begin{equation}
g_{\mathrm{ref}}[k] = \frac{1}{N}\sum_{n=1}^N a_n \cdot f_n[k] \hspace{1cm} a_n = e^{\mathrm{j}2\pi \cdot (x_n l_\mathrm{ref} + y_n m_\mathrm{ref})} 
\end{equation} 
The grid coordinates, for which the beam is scanned, are chosen as:
\begin{eqnarray}\label{eq:grid}
l_p &= p\cdot \Delta l & p \in \{-M/2,\cdots, M/2\} \nonumber \\
m_q &= q\cdot \Delta m & q \in \{-M/2,\cdots, M/2\} 
\end{eqnarray}
The spacings in the grid, $\Delta l$ and $\Delta m$, and number of points per dimension, $M+1$, determine the size and resolution of the aperture image. We discuss the choice of these values below.
The scan signals, $s_{pq}[k]$, are calculated for all scanning directions centered around the reference direction:
\begin{equation}
s_{pq}[k] = \frac{1}{N}\sum_{n=1}^N a_n e^{\mathrm{j}2\pi \cdot (x_n l_p + y_n m_q)}\cdot f_n[k]
\end{equation} 
Then, a quantity proportional to the sampled array factor of the TUT is computed by cross correlating the reference signal with the scan signal for each grid point. The ``Beam-Correlation'' is defined as
\begin{equation}
A[l_p,m_q] = \frac{1}{K}\sum_{k=1}^K g^*_\mathrm{ref}[k] \cdot s_{pq}[k].
\end{equation}
The two-dimensional Fast Fourier Transformation (FFT) of the Beam-Correlation provides the aperture image (\ref{eq:fourier}).

\subsection{Fourier Relationship}
A fundamental principle of the discrete Fourier Transformation is that the sampling interval in one domain is reciprocal to the range of covered values in the other domain. In this case, the choice $\Delta l$ and $\Delta m$ determines the area that is covered by the aperture image. The critical angular sampling interval that is necessary to get an image of the whole aperture is set by the size of the aperture and the observation wavelength by the following relationship, where $D_\mathrm{East}$ and $D_\mathrm{North}$ depict the diameter of the array:
\begin{equation}\label{eq:deltalm}
\Delta l_\mathrm{max} = \frac{\lambda}{D_\mathrm{East}} \hspace{1cm} \Delta m_\mathrm{max} = \frac{\lambda}{D_\mathrm{North}} 
\end{equation}
According to (\ref{eq:grid}), the number of sampling points in each angular dimension is $M+1$. With $\Delta l$ and $\Delta m$ being fixed by the size of the aperture, the choice of $M$ determines for what range of angles the array factor is scanned. Also, the number of points in each dimension of the aperture image is $M+1$. Thus, computing $A[l_p,m_q]$ for a wide range of angles increases the resolution of the aperture image.
We note that $A[l_p,m_q]$ can be computed for all $(l_p,m_q)$, including for values $> 1$, not just for combinations that map to the celestial sphere. By sampling the $(l,m)$ grid for values $> 1$, the {$(x,y)$} grid resolution can be resolved as required.

\section{Measurements on the EDA-2}
\label{sec:measurements}
The Engineering Development Array 2 is located at the Murchison Radio-astronomy Observatory (MRO) in Western Australia. It consists of 256 dual polarisation dipole antennas, pseudo-randomly spread on a circular field with a diameter of {35}\,{m}.
The array operates in the frequency range between 50\,MHz and {350}\,{MHz}.
Groups of 16 antennas are connected to an in-field ``SMART-Box'' in which the signals are converted to be transmitted via optical fiber to the control building, where they are converted back to electrical signals, amplified, filtered and digitised \cite{2017JAI.....641014N}.
Signals from each antenna are passed through an oversampled polyphase filterbank that produces ``coarse'' channels of $\approx$\,0.93\,MHz bandwidth \cite{2017JAI.....641015C}. Short dumps of raw voltage data from all antennas for a single coarse channel can be captured and written to disk for offline processing.
The EDA-2 has identical station configuration and type of antenna to the EDA-1 \cite{2017PASA...34...34W}, with the difference that all antennas are individually digitised in EDA-2.

\begin{figure}
    \centering
    \includegraphics[width=0.7\textwidth]{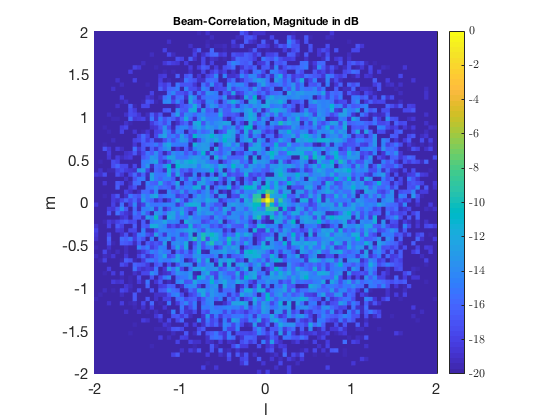}
    \caption{The magnitude of the Beam-Correlation from data at 149\,MHz, in dB. \add[RW]{The coordinates are shown looking ``down'' on the sky.}
    \remove[UK]{The Beam-Correlation is computed on a $81\times 81$-grid and windowed with a Gaussian window.}}\label{fig:beam_mag}
\end{figure}

\begin{figure}
    \centering
    \includegraphics[,clip,width=0.7\textwidth]{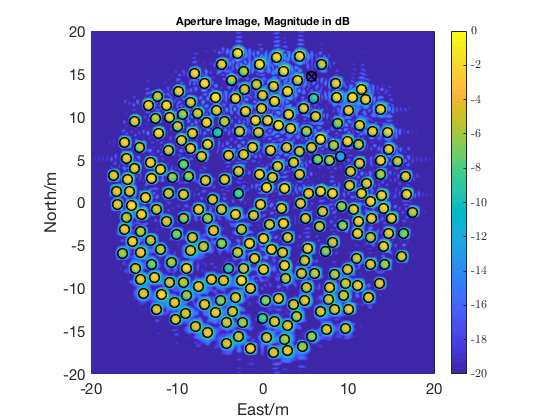}
    \caption{The magnitude of the aperture image \remove[UK]{which is the Fourier Transformation of the Beam Correlation}in dB. Antenna positions are marked with a circle, a previously flagged antenna with a cross.}
    \label{fig:ap_mag}
\end{figure}

\begin{figure}
    \centering
    \includegraphics[width=0.7\textwidth]{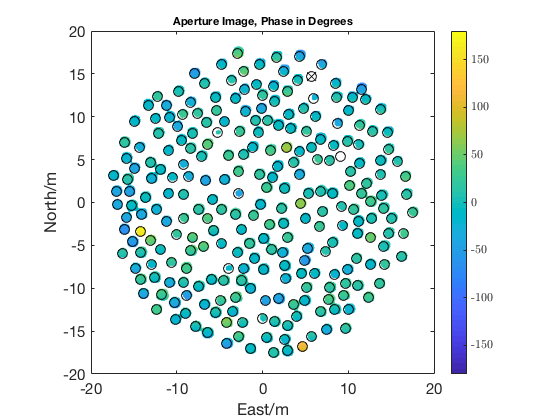}
    \caption{Phase plot of the Aperture Image, in degrees. Patches with a corresponding magnitude of less than -10\,dB are faded. {Antenna positions are marked with a circle, the previously flagged antenna with a cross.}\remove[UK]{For a perfectly calibrated array, the excitation phases of all antennas would be equal.}}
    \label{fig:ap_ph}
\end{figure}

We applied the technique to approximately 0.5\,s of voltage data in a single coarse channel of the EDA-2, captured around noon of
{2020-05-20}.
The data were centred on 149\,MHz.
The Sun was used as strong calibration source, because its angular width is sufficiently small compared to the station beam. The Beam-Correlation is computed for $\Delta l$ and $\Delta m$ chosen according to equation (\ref{eq:deltalm}) and $M = 80$.
Figure \ref{fig:beam_mag} depicts the Beam-Correlation for
{the data}.
To reduce ringing artefacts in the aperture image, we multiply the $81 \times 81$-point Beam-Correlation with a two-dimensional Gaussian window (with $\sigma = 1$ in both $l$ and $m$ dimension). Afterwards, in order to {oversample the aperture image}, the Beam Correlation is zero-padded to a size of $512 \times 512$. Finally, the Fourier Transform {is} computed, which is shown in Figures \ref{fig:ap_mag} and \ref{fig:ap_ph}.

In the aperture images shown in Figures \ref{fig:ap_mag} and \ref{fig:ap_ph}, the previously measured antenna positions are marked with a circle. Any antennas that were previously identified as having poor or no signal, are tagged with a cross, however for the sake of illustration their data have still been included in the process.
Malfunctioning antennas would normally be ``flagged'' and excluded from data processing.
In the phase plot, patches with a corresponding magnitude of less than -10\,dB are faded.
The magnitude of the aperture image corresponds very well with the antenna positions.
As expected, antennas with poor or no signal appear strongly attenuated or not at all in the magnitude plot.


From the phase distribution (Figure \ref{fig:ap_ph}) it is possible to assign excitation phases to all antennas, using their known positions in the aperture. If the array is already calibrated, the phase excitation should be equal for all antennas. Without initial phase calibration, the excitation phases from the image provide the receive path phases of all antennas.

\begin{figure}
    \centering
    \includegraphics[width=0.7\textwidth]{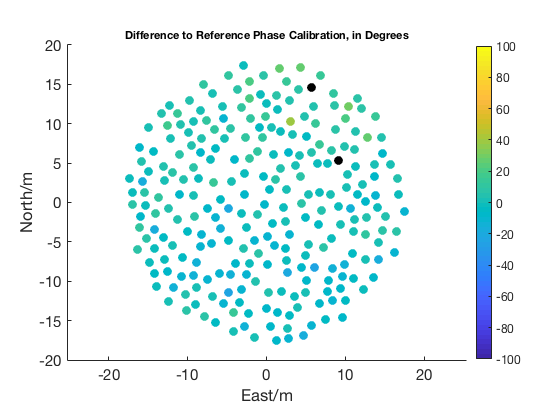}
    \caption{The difference in the phase solution of holographic and reference calibration, in degrees. Flagged and low gain antennas are marked in black. \remove[UK]{Considering all functioning antennas, the RMS phase difference is 9.2$^{\circ}$.}
    }
    \label{fig:phase_diff}
\end{figure}

To evaluate their validity, they are compared with a reference calibration solution, which was measured by a more conventional method. Visibilities were computed between all antenna pairs then phased towards the Sun.
A standard radio astronomy calibration procedure that assumes a compact source at the phase centre was used to solve for antenna-based complex gains. Baselines shorter than $5 \lambda$ were not used.
For each element, the difference between both calibration phases is depicted in Figure \ref{fig:phase_diff}
.

Flagged antennas and antennas with a magnitude of less than {-10}\,{dB} (according to Figure \ref{fig:ap_mag}) are marked in black.
The RMS phase difference between the conventional calibration solutions and the aperture phases was measured to be $9.2^{\circ}$, excluding the flagged antennas.

\begin{figure}
    \centering
    \includegraphics[width=0.7\textwidth]{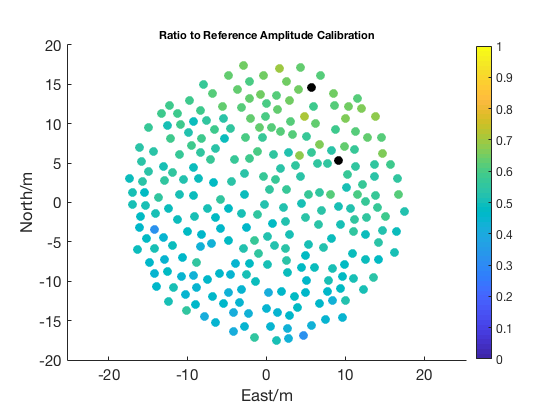}
    \caption{Ratio of the measured gain amplitude between holographic and conventional calibration methods.
    }
    \label{fig:gain_ratio}
\end{figure}

Fig \ref{fig:gain_ratio} shows the ratio of amplitude of the calibration gain between the holographic and conventional calibration methods. The overall scaling is arbitrary since the holographic method does not depend on the absolute flux scale. While some systematic variation is present (a small gradient in the gain amplitude across the array), it is clear that the methods produce similar gain amplitude estimates up to a scaling constant.

\begin{figure}
    \centering
    \includegraphics[width=0.7\textwidth]{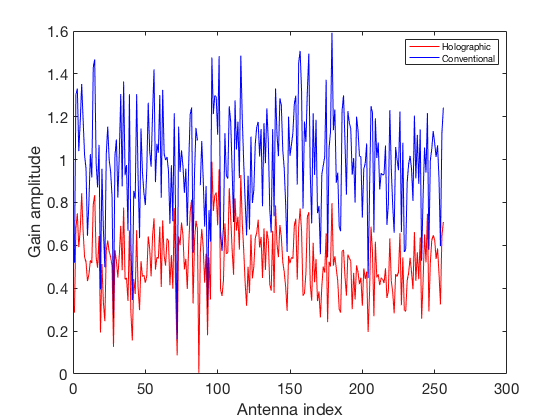}
    \caption{The measured gain amplitude of the holographic and conventional calibration methods, as a function of antenna index. The blue line shows the gain amplitude from conventional calibration, the red line from holographic calibration. There is substatial scatter of the gain amplitudes between antennas. 
    }
    \label{fig:gain_vs_ant}
\end{figure}

Fig \ref{fig:gain_vs_ant} \add[RW]{shows the gain amplitude as a function of antenna index for the two methods. Here the mapping from antennna index to antenna location is not relevant; the point of the figure is that it shows substantial variation in amplitude between antennas. However, the overall pattern is the same for both methods and results in the largely uniform gain ratio shown in Fig} \ref{fig:gain_ratio}. \add[RW]{We discuss the small gradient in amplitude below.}

We do not expect the conventional calibration and holographic methods to produce exactly the same result. With conventional calibration, baselines shorter than 5$\lambda$ were discarded so that the solution was not biased by large-scale emission in the sky, whereas the holographic method forms station beams where all antennas are included.
A detailed treatment of the errors (especially systematic) is beyond the scope of this paper, however we would expect that since the station beam cannot ``discard short baselines'' this holographic technique will be more susceptible to systematic errors due to all the other sources (and large-scale emission) in the sky besides the reference source.

\begin{figure}
    \centering
    \includegraphics[width=0.7\textwidth]{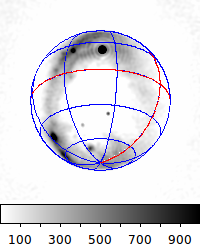}
    \caption{The image of the sky made from the dataset used in this work. The greyscale is in Jy/beam, and has been ``burned out'' to highlight weak/faint emission. For reference, the Sun (top center) was calibrated as a 51,000\,Jy source.
    }
    \label{fig:sky}
\end{figure}

To illustrate this point, Fig \ref{fig:sky} \add[RW]{shows the sky imaged from the data used in this work. The color scale has been deliberately adjusted to highlight weak/faint emission compared to the very bright Sun. Emission from the Galactic plane, Galactic supernova remannts, and some brighter compact sources is apparent.}

A related potential bias of the method is due to the autocorrelation of system noise (not signals from the sky) of the reference beam with the steered beam. Because phases are applied to each antenna's signal to steer the beam, the system noise is effectively re-randomised for each unique pointing direction, and decorrelates on the angular size scale of the station beam. In this situation, the signal in the reference beam is dominated by signals from the sky: the Sun is a $\sim 50$\,kJy source and the array SEFD is approximately 1800\,Jy, however the method may be biased in the case of a lower signal-to-noise ratio from the reference source.

\add[RW]{The bias caused by self-noise and large-scale structure discussed here can of course be eliminated by using a reference beam from a different antenna or station, as is done in conventional dish-based holography. In the context of SKA-Low, this should be possible for stations in close proximity by using an existing calibrated station to generate a reference beam.
For this work the focus is to demonstrate the ability and utility of phased arrays to self-generate a reference beam for the holographic process.}

\begin{figure}
    \centering
    \includegraphics[width=0.7\textwidth]{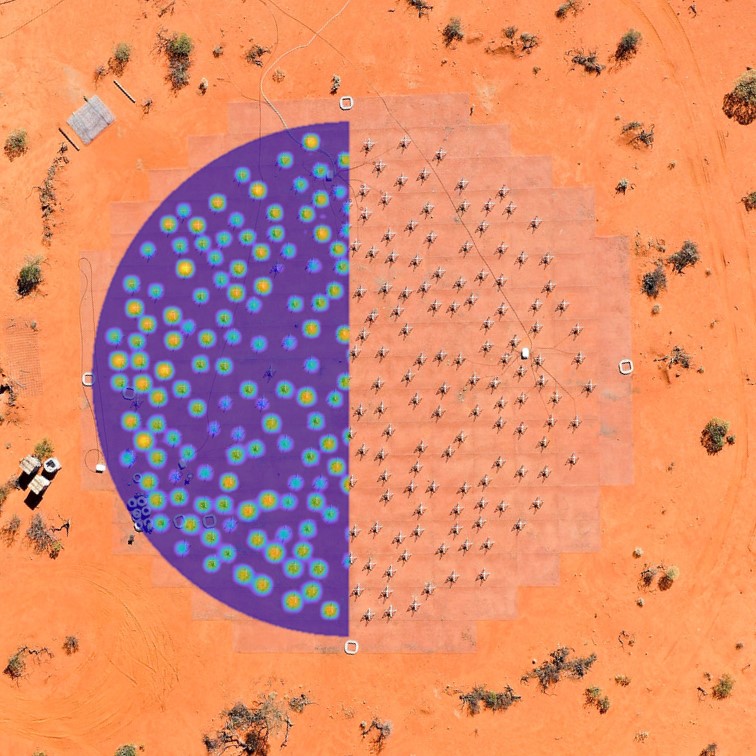}
    \caption{Overlay of {an} aperture image with aerial drone photo of the EDA-2. The aperture image is partially transparent, and antenna dipoles can be seen in the locations of the peaks in the aperture as expected.}
    \label{fig:aerial_overlay}
\end{figure}

\section{Conclusion}
Holographic aperture imaging is mainly associated with dish antennas although the measurement procedure for phased array telescopes with multi-beam capabilities is much simpler. The holographic calibration technique suggested here was successfully tested on a phased array telescope, which is similar in its layout to the envisaged SKA-Low stations. 

With less than one second's worth of voltage data, the technique provides the ability to:
\begin{itemize}
    \item measure the relative complex gain of each antenna in the apeture array
    \item verify {antenna} mappings through the analogue and digital signal paths (see Figure \ref{fig:aerial_overlay})
    \item through regular measurements, monitor the phase or delay of each antenna as a function of time, which can be used to detect changes of delays in the system
\end{itemize}

Hence, the aperture image provided by this method is a simple tool that can be useful for calibration and maintenance of a phased array telescope.

{Finally, the calibration solution from holography takes both the receive path gains and the EEPs for the pointing direction of the calibrator into account. Thereby the receive path gain factor is direction independent, while the contribution of the EEPs depends on the celestial position of the calibration source.}
{Similarly to conventional calibration, the calibration solution from holography incorporates contributions from both EEPs and direction independent receive path gains.}
By applying the technique with calibration sources at various sky-coordinates and comparing the solutions, it might be possible to extract EEPs for all individual antennas and use them to verify simulated EEPs. However, such exact measurements might require a more complex measurement, such as using a separate antenna to form an independent reference beam.

Code and data used for this project is available at \url{https://doi.org/10.25917/5f35032844ed8}.

\acknowledgments
This scientific work makes use of the Murchison Radio-astronomy Observatory, operated by CSIRO. We acknowledge the Wajarri Yamatji people as the traditional owners of the Observatory site.
The acquisition system was designed and purchased by INAF/Oxford University and the RX chain was design by INAF, as part of the SKA design and prototyping program.

We thank the anonymous referees for their insightful comments and patience while the manuscript was updated.

%
%
\newcommand{\aj}{AJ}
\newcommand{\pasa}{PASA}
\newcommand{\aap}{A\&A}
\newcommand{\pasp}{PASP}
\newcommand{\procspie}{SPIE}
\bibliography{citations.bib}

%
%
%
%
%

\end{document}